# MESSAGE EMBEDDED CIPHER
# USING 2-D CHAOTIC MAP

Mina Mishra[1] and Dr. V.H. Mankar[2]

[1]Ph. D. Scholar, Department of Electronics & Telecommunication, Nagpur University,
Nagpur, Maharashtra, India
minamishraetc@gmail.com

[2]Senior Faculty, Department of Electronics Engineering, Government Polytechnic,
Nagpur, Maharashtra, India
vhmankar@gmail.com

## ABSTRACT

*This paper constructs two encryption methods using 2-D chaotic maps, Duffings and Arnold's cat maps respectively. Both of the methods are designed using message embedded scheme and are analyzed for their validity, for plaintext sensitivity, key sensitivity, known plaintext and brute-force attacks. Due to the less key space generally many chaotic cryptosystem developed are found to be weak against Brute force attack which is an essential issue to be solved. For this issue, concept of identifiability proved to be a necessary condition to be fulfilled by the designed chaotic cipher to resist brute force attack, which is a basic attack. As 2-D chaotic maps provide more key space than 1-D maps thus they are considered to be more suitable. This work is accompanied with analysis results obtained from these developed cipher. Moreover, identifiable keys are searched for different input texts at various key values.*

*The methods are found to have good key sensitivity and possess identifiable keys thus concluding that they can resist linear attacks and brute-force attacks.*

## 1. INTRODUCTION

For last several years many efforts have been made to use chaotic systems for enhancing some features of communications systems. Chaotic signals are highly unpredictable and random-like nature, which is the most attractive feature of deterministic chaotic systems that may lead to novel (engineering) applications. Some of the common features between chaos and cryptography are being sensitivity to variables and parameters changes. An important difference between chaos and cryptography lies on the fact that systems used in chaos are defined only on real numbers, while cryptography deals with systems defined on finite number of integers. Nevertheless, we believe that the two disciplines can benefit from each other. Thus, for example, as it is shown in this paper, new encryption algorithms can be derived from chaotic systems. On the other hand, chaos theory may also benefit from cryptography: new quantities and techniques for chaos analysis may be developed from cryptography.

During the past two decades, there has been tremendous interest worldwide in the possibility of using chaos in communication systems. Many different chaos-based decryption algorithms have been proposed up to date.

The aim of this paper is to construct and crypt analyze two of the stream symmetric chaotic ciphers constructed using one of the latest chaotic scheme known as message-embedded scheme [6] [7]. Both of the developed methods use 2-D chaotic maps, Duffings and Arnolds Cat map.

Parameters of the respective chaotic maps act as secret key in the ciphers due to which complexity and key space is increased compared to 1-D chaotic map. Both of the ciphers are analyzed for key space, avalanche effect and strength against Brute-force and Known-plaintext attack.

A cryptanalytic method based on the identifiability concept, solves the problem of less key space in chaotic ciphers. It is possible to test about the cipher strength against Brute-force attack using it. Both of the mentioned ciphers are concluded to provide security against the Brute-force attack. Identifiability concept fulfills the necessary condition but not sufficient as the developed cryptosystems must be tested for sensitivity and other statistical tests to result in a robust cipher. Thus both the ciphers are tested for sensitivity and it is concluded that some of the keys selected from domain of key space of the ciphers seem to have good key sensitivity and resist known plaintext attack for the available first two characters of plaintext.

This paper is organized into five sections as follows. Section II, presents the background and in section III algorithm for encryption used in developing ciphers is provided. Then in section IV, analysis result in tabulated form and discussions are presented. Section V, discusses about the conclusions derived.

## 2. BACKGROUND

Message-Embedded Scheme: According to this scheme at the transmitter side, the plain text is encrypted by an encryption rule which uses non-linear function and the state generated by the chaotic system in the transmitter. The scrambled output signal is used further to drive the chaotic system such that the chaotic dynamics is changed continuously in a very complex way. Then another state variable of the chaotic system in the transmitter is transmitted through the channel.

At the receiver side, the reconstruction of the plaintext is done by decrypting the input by using the reverse of encryption method.

This method can be illustrated along with diagram as shown in fig 1.

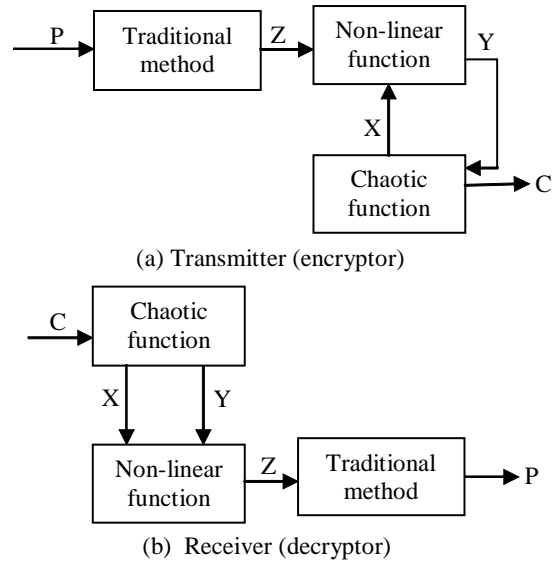

(a) Transmitter (encryptor)

(b) Receiver (decryptor)

Figure 1. Hybrid message-embedded chaotic cryptosystem

P: plaintext; C: cipher text; X: state of chaotic function;
Z &Y: Intermediate encrypted plaintext;

In the present work modified form of maps has been used that is generated and its randomness can be clearly seen in fig 2(a) and (b) and fig 3(a) and (b) respectively.

**Arnold's Cat Map:** Arnolds Cat map is a 2-D discrete-time dynamical system, which takes a point(x, y) in the plane and maps it to a new point using equations:

$$x(k+1) = (a-1) \mod [2x(k) + y(k), N];$$

$$y(k+1) = \mod [x(k) + (1-b) y(k), N];$$

a, b and N are parameters on which the map depends. At a=0.3, b=0.345, map exhibits chaotic nature.

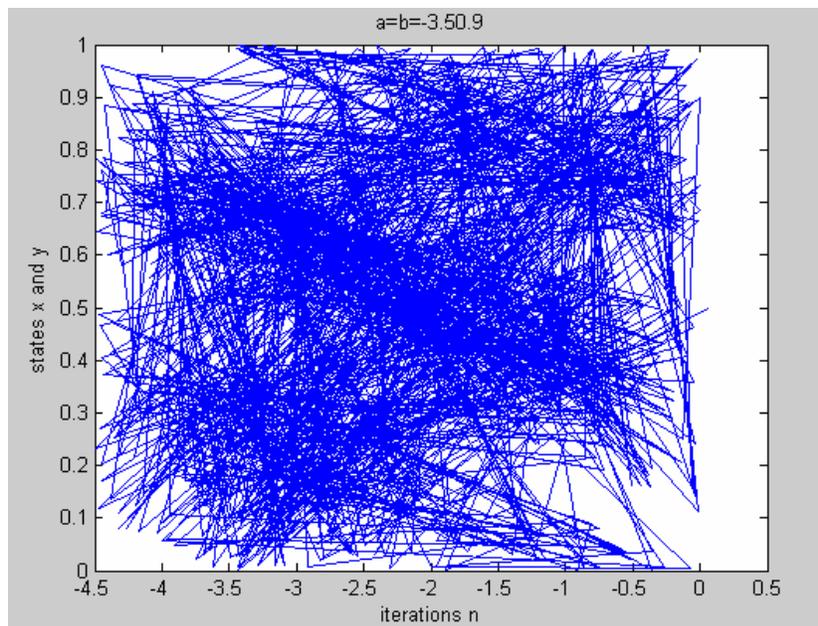

Figure 2. (a)  Plot of Arnolds Cat Map at x(0) = 0.5, y(0) = 0.06, a = - 3.5, b =  0.9, n=1000.

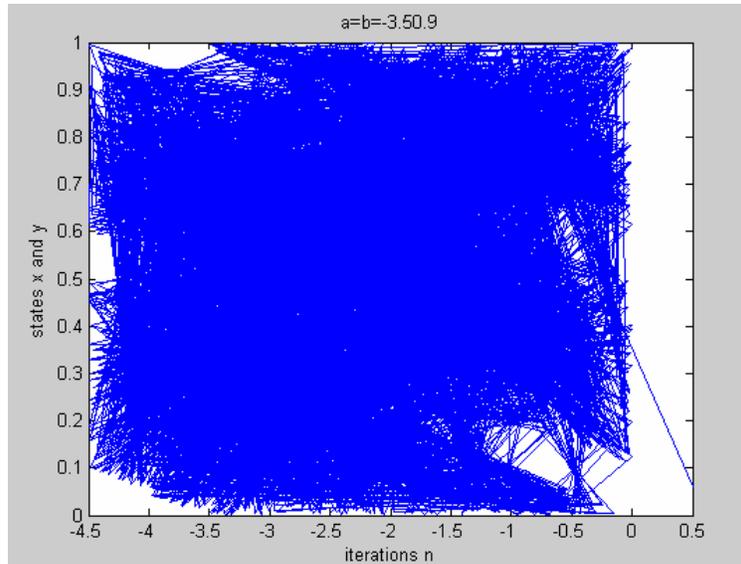

Figure 2. (b) Plot of Arnolds Cat Map at x(0) = 0.5, y(0) = 0.06, a = - 3.5, b = 0.9, n=5000.

**Duffings map:** Duffings map is a 2-D discrete-time dynamical system, which takes a point (x, y) in the plane and maps it to a new point using equations:

$$x(k+1) = y(k);$$

$$y(k+1) = -bx(k) + ay(k) - y(k)^3$$

a and b are parameters on which the map depends. At a = 2.75, b = 0.2, map exhibits chaotic nature.

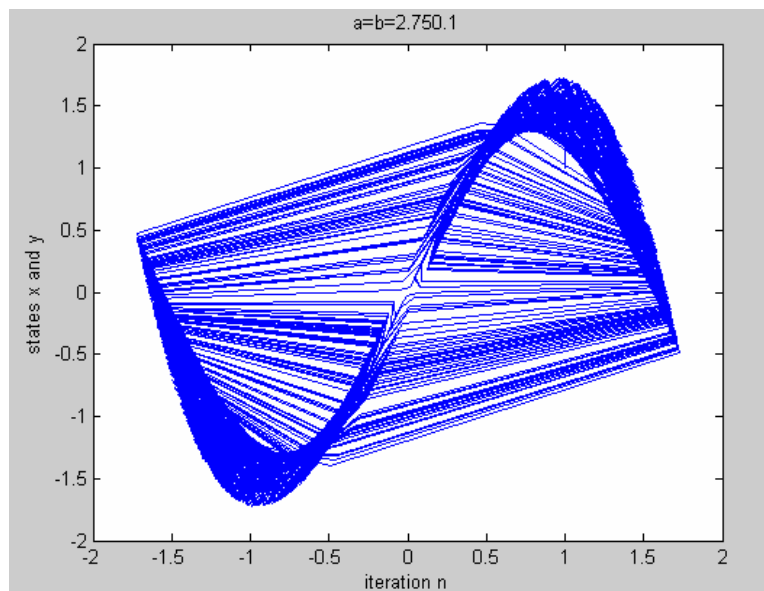

Figure 3. (a) Plot of Duffings Map at x(0) = - 0.04, y(0) = 0.2, a= 2.75, b=0.1, n=1000.

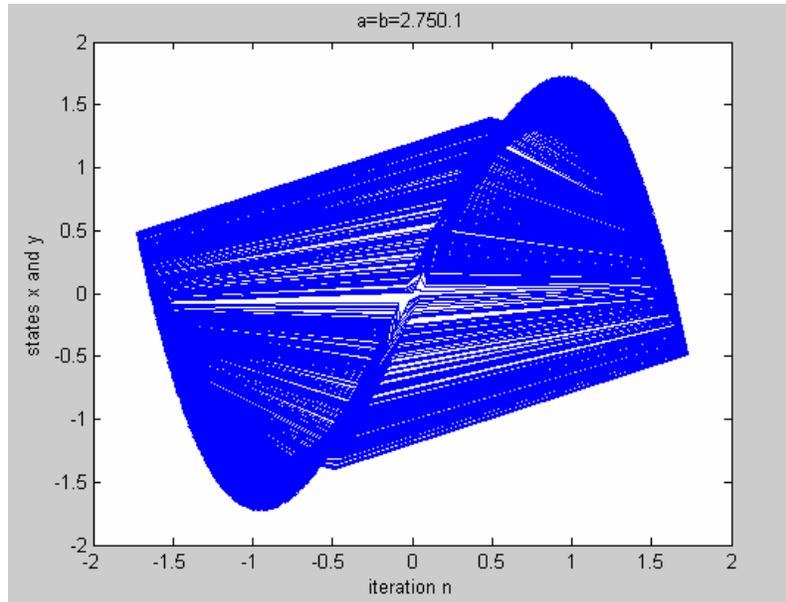

Figure3. (b) Plot of Duffings Map at x(0) = - 0.04, y(0) = 0.2, a= 2.75, b=0.1, n=5000.

**Non-Linear Function:** Modular function is used as non-linear function in the construction of ciphers.

The Mod numeric function returns the remainder when the dividend is divided by the divisor. The result is negative only if the dividend is negative. Both the numbers must be integers. The function returns an integer. If any number is NULL, the result is NULL. For example:

Mod (5, 3) returns 2.

Mod (-5, 3) returns -2.

**Cryptanalysis:** Cryptanalysis is the study of attacks against cryptographic schemes to disclose its possible weakness. During crypt analyzing a ciphering algorithm, the general assumption made is that the cryptanalyst knows exactly the design and working of the cryptosystem under study, i.e., he/she knows everything about the cryptosystem except the secret key. It is possible to differentiate between different levels of attacks on cryptosystems. They are briefly explained as follows:

**1. Cipher text-only attack:** The attacker possesses a string of cipher text.

**2. Known plain text:** The attacker possesses some portion of plain text and the corresponding cipher text.

**3. Chosen plain text:** The attacker has obtained temporary access to the encryption machinery. Hence he/she can choose a plain text string, p, and construct the corresponding cipher text string.

**4. Chosen cipher text:** The attacker has obtained temporary access to the decryption machinery. Hence he/she can choose a cipher text string, c, and construct the corresponding plain text string.

**5. Brute Force Attack:** A brute force attack is the method of breaking a cipher by trying every possible key. The brute force attack is the most expensive one, owing to the exhaustive search.

In addition to the five general attacks described above, there are some other specialized attacks, like, differential and linear attacks.

Differential cryptanalysis is a kind of chosen-plaintext attack aimed at finding the secret key in a cipher. It analyzes the effect of particular differences in chosen plaintext pairs on the differences of the resultant cipher text pairs. These differences can be used to assign probabilities to the possible keys and to locate the most probable key.

Linear cryptanalysis is a type of known-plaintext attack, whose purpose is to construct a linear approximate expression of the cipher under study. It is a method of finding a linear approximation expression or linear path between plaintext and cipher text bits and then extends it to the entire algorithm and finally reaches a linear approximate expression without intermediate value.

**Security Analysis:** Various cryptanalytic procedures are developed to test the validity of newly constructed ciphers and they are as follows:

**(a) Key Space Analysis:** Key space belongs to the chaotic region of the system in case of chaotic ciphers. The total key space is a product of all the parameters involved. Once the key has been defined and key space has been properly characterized, the good key is chosen randomly from the large key domain.

**(b) Identifiability Test method:** From the crypto graphical point of view, the size of the key space should not be smaller than 2100 to provide a high level security so that it can resist all kind of Brute force attack. A fundamental issue of all kinds of cryptosystem is the key. No matter how strong and how well designed the encryption algorithm might be, if the key is poorly chosen or the key space is too small, the cryptosystem will be easily broken. Unfortunately, chaotic cryptosystem has a small key space region and it is non-linear because all the keys are not equally strong. The keys should be chosen from the chaotic region. To solve the problem of small key space and weakness against brute force attack, identifiability concept is quite advantageous.

A cryptanalytic procedure, known as output equality based on the identifiability concept, is carried out on the developed ciphers. It is found that in chaotic ciphers, there exists a unique solution for a particular input for certain domain of values of parameters. The response of any system to a particular input is the solution of that particular system and it contains all the information about the parameters of system. In the discussed ciphers, system parameters are acting as a secret key. This type of analysis is also known as parametric analysis.

The output equality method is explained as follows:

"For the same inputs and initial condition, transmitter system is parameterized at different values of parameter taken from the existing domain of parameter space, if the output response of the system obtained after some value of iteration, parameterized at a particular value coincides with the output response of the same system parameterized at some other value of parameter within the domain for the same number of iteration, then both the parameters are said to be equal and identifiable. The system is said to possess unique solution at that particular value of parameter and the system is said to be structurally identifiable."

If parameter of the transmitter is identifiable, it is more difficult for the eavesdropper to find it by a brute force attack. Consequently, this parameter can play the role of the secret key against brute force attack. If parameter is not identifiable, the eavesdropper has a higher favorable chance to find it by a brute force attack and thus, the parameter vector cannot play the role of the secret key against brute force attack.

**(c) Plaintext sensitivity Test Method:** The percentage of change in bits of cipher text obtained after encryption of plaintext, which is derived by changing single bit from the original plaintext from the bits of cipher text obtained after encryption of original plaintext. With the change in single bit of plaintext, there, must be ideally 50% change in bits of cipher text to resist differential cryptanalysis (chosen-plaintext attack) and statistical analysis.

**(d) Key sensitivity Test Method:** The percentage of change in bits of cipher text obtained after encryption of plaintext using key, which is flipped by single bit from the original key, from bits of cipher text obtained after encryption of plaintext using original key, which requires ideally 50% change in cipher text bits to resist Linear and statistical attacks.

**(e) Known plaintext attack Method:** For observing this attack on developed cryptosystem it is assumed that the opponent knows everything about the algorithm, he/she has the corresponding cipher text of plaintext and some portion of plaintext. With this much information, the opponent tries to find out the secret key.

## 3. ALGORITHM FOR THE DEVELOPED CIPHERS

Encryption Algorithm:

**Step-1:** Read plaintext and key vector.

**Step-2:** Convert plaintext into its ASCII values.

**Step-3:** Each value of ASCII values are transformed using following steps:

(a) The chaotic map is iterated for a number of times to output a random state.

(b) ASCII value is mixed with non-Linear function and the output state of chaotic system obtained after a fixed value of iteration.

(c) Again chaotic system is iterated for a fixed number of times and an output state is obtained.

(d) The response obtained in (b) is mixed with the output state obtained in (d) and output values are obtained as cipher text.

**Step-4:** Convert cipher text into characters.

**Step-5:** Read the cipher text.

Decryption algorithm is reverse of encryption process and the original information is retained using the same secret key using which encryption is being done and it is kept secret between authenticated sender and receiver only.

## 4. RESULTS AND ANALYSIS

The simulated result data produced after analyzing both the ciphers using above discussed (section II) cryptanalytic procedures is summarized with the help of table 1 and 2. Twenty different values of keys are chosen from key space of respective ciphers and are analyzed for its security. From both the observation tables, it can be seen that plaintext sensitivity of Arnold's cat cipher ranges from 0.5 to 2.5 % and Duffings cipher ranges from 0.5 to 2 %, which is not sufficient. Key sensitivity for each of the cipher ranges from 0 to 36 % and from 0 to 51 %, respectively. Thus key sensitivity property of some keys from both the ciphers shows satisfactory values. Both ciphers are robust against known plaintext attack for the available first two characters of plaintext. Key space of ciphers shows lesser range than compared to the required limit i.e. 2100 to resist Brute-force attack but identifiable keys conclude that the developed ciphers can resist Brute-force attack.

**a. Arnold's Cat cipher:** Key space is from [-5 0.4] to [-0.9 1.5] = 5 x $10^{16}$

Table 1. Analysis Table for Arnold's Cat

| Sl. No. | Plain text | Key value | Ciphertext | Plaintext sensitivity (in %) | Key sensitivity (in %) | Domain for key With increment = 0.0001 | Identifiability of key for iteration value =2 or 3 | Robustness against known plaintext attack for p=[p1 p2]. | Whether key can act as secret key against Brute Force attack? |
|---|---|---|---|---|---|---|---|---|---|
| 1. | What is your name? | [0.0034 0.0013] | k¢³¬¿k´¾kÅ°À½k¹¬° | 1.9737 | 24.3421 | [0 0]to[0.005 0.004] | I | R | YES |
| 2. | I am going to market . | -4.4977 0.2034] | kk¬˛k²º'¹²k¿°k˛¬½²¶°¿y | 1.7045 | 21.5909 | [-4.5 0.2]to[-4.495 0.204] | I | R | YES |
| 3. | My college name is s.s.c.e.t . | -2.9982 0.6981] | kÅk®º�··º²°k¹¬°k´¾k ¾y¾y®y°y¿y | 1.2500 | 25.4167 | [-3 -0.7]to[-2.995 -0.696] | I | R | YES |
| 4. | Hello! how are you? | [-2.992 -0.694] | i®µµˍj±˛Ài²»®iÅˍ¾ | 1.3158 | 0 | [-2.995 -0.696]to[-2.99 -0.692] | NI | R | NO |
| 5. | Sita is singing very well. | [-2.99 - 0.692] | h±¼©h±»h»±¶¬¶±¶ h¾ °Áh¿´´v | 0.4630 | 36.1111 | [-2.99 - 0.692]to[-2.985 -0.688] | NI | R | NO |
| 6. | Ram scored 98 marks in Maths. | [-2.985 -0.688] | lµÀ- lµ¿l¿µº³µº¶lÁ±¾ÁlÃ±ˍˍz | 0.4630 | 24.0741 | [-2.985 - 0.688] to[-2.98 -0.684] | NI | R | NO |
| 7. | Jaycee publica tion. | -0.9957 0.0101] | lÀ¯±±l¼Áˍµ¯- Àµ»ºzU | 0.5952 | 23.8095 | [-1 0.01]to[-0.995 0.014] | I | R | YES |
| 8. | Thank you,sir . | [-0.99 0.018] | k³¬¶kÀ°Àw¾¼¶y | 2.5000 | 26.6667 | [-0.995 0.014]to[-0.99 0.018] | NI | R | NO |
| 9. | The match was very excitin g. | -0.9853 0.0213] | k³°k¬¬¿¿®³kÀ¬¾kÅ¿°½ Åk¬Ã®¿´¬¹²y | 1.2931 | 0 | [-0.99 0.018]to[-0.985 0.022] | I | R | YES |
| 10. | I will be leaving at 9p.m. | [0.2035 1.4009] | llÀµˍˍl®±l˛Àµº³l- Àl¼¼z¹z | 0.4630 | 28.2407 | [0.2 1.4]to[0.205 1.404] | I | R | YES |
| 11. | How are you? | -5 0.4 | °Åk¬½°kÀ°Å | 3.1250 | 31.25 | [-5.0 0.4] to [-4.995 0.4031] | NI | R | NO |
| 12. | Meet me after 5p.m. | [-4 0.5] | ¯¯¾¼jˍ j«°¾¯¼¼j °x˝x | 2.6316 | 38.1579 | [-4.0 0.5 ] to [-3.995 0.5031 ] | NI | NR | NO |

| 13. | I have a gift for you. | [-3 0.6] | h°¾-h©h¯±®¼h® ·°hÁ·½v | 1.1364 | 24.4318 | [-3. 00 0.6 ] to [-2.995 0.6031 ] | I | NR | YES |
|---|---|---|---|---|---|---|---|---|---|
| 14. | We will go for walk. | [-2 0.7] | l£±lÃµ‚‚l³»l²»¾lÃ- ‚·z | 0.5952 | 31.5476 | [-2 0.7] to [ -1.995 0.0.7031] | NI | NR | NO |
| 15. | Study different papers. | [-1 0.8] | ¿À¯Äk¯´±±°½°¹¿k»¬»°½½¾y | 0.5435 | 31.2500 | [-1 0.8] to [-0.995 0.8031] | I | R | YES |
| 16. | How to do analysis? | [-0.9 0.9] | 1 »ÃlÀ»l°»l-°-‚Ã¿µ¿ | 0.6250 | 27.5000 | [-0.9 0.9] to [-0.895 0.9031] | I | R | YES |
| 17. | Hai! Where are you going? | [-4.99 1.0] | 1 -µml£´±¾±l-¾±lÀ»Ál³»µ°³ | 0.4808 | 20.6731 | [- 4.99 1.0] to [-4.98 1.01] | NI | NR | NO |
| 18. | Dolly, are you coming with me? | [-3.78 1.1] | j ¹¶¶Ãvj«¼¯jÃ¹¿j-¹·³‚±jÃ³¾²j·¯ | 0.4032 | 0.0 | [-3.78 1.1 ] to [-3.77 1.1 ] | NI | R | NO |
| 19. | Children are playing in park. | [-2.2 1.2] | j ²³¶®¼¯ j«¼¯j¶«Ã³‚±j³ j°«¼µxj | 0.4032 | 24.5968 | [ -2.2 1.2] to [-2.2 1.3 ] | NI | NR | NO |
| 20. | I shall go to cinema. | [-2.5 1.4] | h h»°©´h¯·h¼·h«±¶-µ©v | 0.5682 | 19.8864 | [- 2.5 1.4] to [ -2.5 1.5] | NI | NR | NO |

NI – Non – Identifiable; I – Identifiable; R- Robust; p[p1 p2…p n]- First 'n' characters of available plaintext string. ]; NR – Not robust;

**b. Duffings cipher:** Key space is from [1.8 -0.59] to [2.9 0.2] = 9 x $10^{14}$

Table 2. Analysis Table for Duffings.

| Sl. No. | Plaintext | Key value | Ciphertext | Plaintext sensitivity (in %) | Key sensitivity (in %) | Domain for key With increment = 0.0001 | Identifiability of key for iteration value =2 or 3 | Robustness against known plaintext attack for p=[p1 p2]. | Whether key can act as secret key against Brute Force attack? |
|---|---|---|---|---|---|---|---|---|---|
| 1. | What is your name? | [1.803 -0.584] | k¢³¬¿k¯¾kÃ°À½k¹¬°‚° | 1.9737 | 28.9474 | [1.8 -0.59]to[1.805 -0.586] | NI | R | NO |
| 2. | I am going to market. | [1.8995 0.0068] | !J!bn!hpjoh!up!nbslfu/ | 1.2987 | 0 | [1.895 0.006]to[1.9 0.01] | I | R | YES |
| 3. | My college name is s.s.c.e.t. | [1.9015 0.0100] | !Nz!dpmmfhf!obnf!jt!!t/t/d/f/u/ | 0.9524 | 0 | [1.897 0.008]to[1.902 0.012] | I | R | YES |

| | | | | | | | | |
|---|---|---|---|---|---|---|---|---|
| 4. | Hello!how are you? | [1.8127 - 0.5804] | !Ifmmp"ipx!bsf!zpv @ | 1.5038 | 21.0526 | [1.81 - 0.582]to[1.8 15 -0.578] | I | R | YES |
| 5. | Sita is singing very well. | [1.7547 0.1521] | !Tjub!jt!tjohjoh!wfs z!xfmm/ | 1.0582 | 0 | [1.75 0.15]to[1.75 5 0.154] | I | R | YES |
| 6. | Ram scored 98 marks in Maths. | [1.4045 0.0020] | !Sbn!tdpsfe!:9!nbslt !jo!Nbuit/ | 0.9524 | 0 | [1.4 2.15]to[1.40 5 2.154] | I | R | YES |
| 7. | Jaycee publication. | [1.7544 0.1540] | !Kbzdff!qvcmjdbujp o/ | 1.4286 | 0 | [1.75 0.15]to[1.75 5 0.154] | I | R | YES |
| 8. | Thank you,sir. | [1.6247 0.0937] | !Uibol!zpv-tjs/ | 1.9048 | 0 | [1.62 0.09]to[1.62 5 0.094] | I | R | YES |
| 9. | The match was very exciting. | [2.802 - 0.094] | "Vjg"ocvej"ycu"xgt {"gzekvkpi0 | 0.4926 | 51.2931 | [2.8 - 0.1]to[2.805 -0.096] | NI | R | NO |
| 10. | I will be leaving at 9p.m. | [2.0045 0.1509] | !J!xjmm!cf!mfbwjo h!bu!:q/n/ | 1.0582 | 0 | [2 0.15]to[2.00 5 0.154] | I | R | YES |
| 11. | How are you? | [1.81 -0.44 | Ipx!bsf!zpv@ | 2.3810 | 0 | [1.81 -0.44] to [1.81 - 0.43] | NI | R | NO |
| 12. | Meet me after 5p.m. | [1.82 -0.57 | !Nffu!nf!bgufs!6q/n / | 1.4286 | 48.7500 | [1.82 -0.57] to [1.83 - 0.57] | NI | R | NO |
| 13. | I have a gift for you. | [1.85 -0.47 | J!ibwf!b!hjgu!gps!z pv/ | 0.6494 | 50 | [ 1.85 - 0.47] to [1.85 - 0.46] | NI | NR | NO |
| 14. | We will go for walk. | [ 1.89 -0.3 | "Yg"yknn"iq"hqt"yc nm0 | 0.6803 | 0 | [ 1.89 -0.3] to [1.9 -0.3] | NI | NR | NO |
| 15. | Study different papers. | [1.9 0.2] | !Tuvez!ejggfsfou!qb qfst/ | 1.1905 | 0 | [1.9 0.2] to [1.7 0.2] | NI | R | NO |
| 16. | How to do analysis? | [2.01 0.11 | !Ipx!up!ep!bobmztjt @ | 1.4286 | 0 | [2.01 0.11] to [2.02 0.11] | NI | R | NO |
| 17. | Hai! Where are you going? | [ 2.05 0.14 | !Ibj"!Xifsf!bsf!zpv! hpjph@ | 1.0989 | 0 | [2.03 0.14 ] to [ 2.05 0.14] | NI | NR | NO |
| 18. | Dolly, are you coming with me? | [ 2.3 0.15 | "Fqnn{."ctg"{qw"eq okpi"ykvj"ogA | 0.4608 | 0 | [ 2. 2 0.15] to [ 2.3 0.15] | NI | NR | NO |
| 19. | Children are playing in park. | [2.5 0.18] | !Dijmesfo!bsf!qmbz joh!jo!qbsl/! | 0.9217 | 0 | [2.5 0.18] to [ 2.6 0.18] | I | NR | YES |
| 20. | I shall go to cinema. | [2.6 0.19] | J!tibmm!hp!up!djof nb/ | 1.2987 | 0 | [2.5 0.19] to [2.6 0.19] | I | NR | YES |

It is seen from the observation table that each of the key possesses individual virtues to resist the respective attacks.

## 5. CONCLUSIONS

This paper presents two encryption methods using Duffings and Arnold's cat maps designed using message embedded scheme and are analyzed for their validity, for plaintext sensitivity, key sensitivity, known plaintext and brute-force attacks. As 2-D chaotic maps provide more key space than 1-D maps thus they are considered to be more suitable. Less key space of the developed ciphers concludes that they cannot resist brute force attack which is an essential issue to be solved. For this issue, concept of identifiability proved to be a necessary condition to be fulfilled by the designed chaotic cipher to resist brute force attack.

The methods are found to have good key sensitivity and possess identifiable keys thus concluding that they can resist linear attacks and brute-force attacks. A comparison table no.3 shows that both ciphers are found to resist Brute-force attack as they consist of identifiable keys. Key sensitivity property is also good for some of the keys selected from domain of key space. Ciphers are determined to resist known plaintext attack for available first two characters of plaintext. If available characters are not the starting characters of plaintext then ciphers shows robustness against the attack for available any number of plaintext characters.

Table 3: Comparison between the two ciphers

| Name of Cipher | key Space | Range of plaintext sensitivity | Range of key sensitivity | Identifiable key | Robust against known plaintext attack | Whether key space $> 2^{100}$ |
|---|---|---|---|---|---|---|
| Duffing's | $9 \times 10^{14}$ | 0.5 to 2 % | 0 to 51 % | Yes | Yes | No |
| Arnold's Cat map | $5 \times 10^{16}$ | 0.5 to 2.5 % | 0 to 36 % | Yes | Yes | No |

## ACKNOWLEDGEMENTS

The authors would like to thank the anonymous reviewers for their valuable suggestions and the proposed references.

## Authors

Mina Mishra, is pursuing Ph.D. (Engg) from Nagpur University, Maharashtra, India. She received M.E. degree specialization in communication in the year 2010. Her research area covers chaotic systems, chaotic cryptology, network security and secure communication.

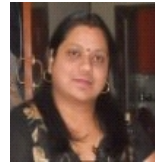

Vijay H. Mankar received M. Tech. degree in Electronics Engineering from VNIT, Nagpur University, India in 1995 and Ph.D. (Engg) from Jadavpur University, Kolkata, India in 2009 respectively. He has more than 16 years of teaching experience and presently working as a Lecturer (Selection Grade) in Government Polytechnic, Nagpur (MS), India. He has published more than 30 research papers in international conference and journals. His field of interest includes digital image processing, data hiding and watermarking.

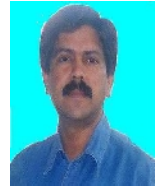